\documentclass[useAMS,usenatbib]{mn2e}
\usepackage{amsmath}
\usepackage{amssymb}
\usepackage{epsfig}
\usepackage{natbib}
\title[The masses of supermassive black holes]{Dark matter haloes determine the masses of supermassive black holes}

\author[C. M. Booth \& J. Schaye]{C. M. Booth$^{1}$\thanks{E-mail: booth@strw.leidenuniv.nl (CMB)} and Joop Schaye$^{1}$\\
$^{1}$Leiden Observatory, Leiden University, PO Box 9513, 2300 RA Leiden, the Netherlands}

\newcommand{\epsf}{\epsilon_{\rm f}}
\newcommand{\epsr}{\epsilon_{\rm r}}
\newcommand{\mbh}{m_{\rm BH}}
\newcommand{\mhalo}{m_{\rm halo}}
\newcommand{\ms}{m_\ast}

\newcommand{\msun}{{\rm M}_{\odot}}
\newcommand{\mseed}{m_{\rm seed}}
\newcommand{\msr}{M_{\rm sr}}
\newcommand{\re}{r_{\rm ej}}

\begin{document}
\pagerange{\pageref{firstpage}--\pageref{lastpage}} \pubyear{2009}
\maketitle
\label{firstpage}
\begin{abstract}
The energy and momentum deposited by the radiation from accretion
flows onto the supermassive black holes (BHs) that reside at the
centres of virtually all galaxies can halt or even reverse gas inflow,
providing a natural mechanism for supermassive BHs to regulate their
growth and to couple their properties to those of their host galaxies.
However, it remains unclear whether this self-regulation occurs on the
scale at which the BH is gravitationally dominant, on that of the
stellar bulge, the galaxy, or that of the entire dark matter halo. To
answer this question, we use self-consistent simulations of the
co-evolution of the BH and galaxy populations that reproduce the
observed correlations between the masses of the BHs and the properties
of their host galaxies. We first confirm unambiguously that the BHs
regulate their growth: the amount of energy that the BHs inject into
their surroundings remains unchanged when the fraction of the accreted
rest mass energy that is injected, is varied by four orders of
magnitude. The BHs simply adjust their masses so as to inject the same
amount of energy. We then use simulations with artificially reduced
star formation rates to demonstrate explicitly that BH mass is not set
by the stellar mass. Instead, we find that it is determined by the
mass of the dark matter halo with a secondary dependence on the halo
concentration, of the form that would be expected if the halo binding
energy were the fundamental property that controls the mass of the BH.
We predict that the black hole mass, $\mbh$, scales with halo mass as
$\mbh \propto \mhalo^\alpha$, with $\alpha \approx 1.55 \pm 0.05$ and
that the scatter around the mean relation in part reflects the scatter
in the halo concentration-mass relation.
\end{abstract}

\section{Introduction}
Almost all massive galaxies are thought to contain a central
supermassive black hole (BH) and the properties of these BHs are
tightly correlated with those of the galaxies in which they reside
\citep[e.g.][]{magg98,ferr00,gebh00,trem02,hari04,hopk07b,ho08}.  It
is known that most of the mass of the BHs is assembled via luminous
accretion of matter \citep{solt82}.  The energy emitted by this
process provides a natural mechanism by which BHs can couple their
properties to those of their host galaxies.  Analytic
\citep[e.g.][]{silk98,haeh98,fabi99,adam01,king03,wyit03,murr05,merl08},
semi-analytic \citep[e.g.][]{kauf00,catt01,gran04,bowe06} and
hydrodynamical
\citep[e.g.][]{spri05,dima05,robe06,sija07,hopk07,dima08,okam08,boot09}
studies have used this coupling between the energy emitted by luminous
accretion and the gas local to the BH to investigate the origin of the
observed correlation between BH and galaxy properties, and the buildup
of the supermassive BH population.

BHs are expected to regulate the rate at which they accrete gas down
to the scale on which they are gravitationally dominant. For example,
gas flowing in through an accretion disk can become so hot that its
thermal emission becomes energetically important. Scattering of the
photons emitted by the accreting matter by free electrons gives rise
to the so-called Eddington limit. If the accretion rate exceeds this
limit, which is inversely proportional to the assumed radiative
efficiency of the accretion disk, then the radiative force exceeds the
gravitational attraction of the BH and the inflow is quenched, at
least within the region that is optically thin to the radiation.

However, observations indicate that the time-averaged accretion rate
is far below Eddington \citep{koll06}, suggesting the presence of
processes acting on larger scales. Indeed, the existence of tight
correlations between the mass of the BH and the properties of the
stellar bulge indicates that self-regulation may happen on the scale
of the bulge \citep[$\sim1$~kpc; ][]{adam01,hopk07}, far exceeding the
radius within which the BH is gravitationally dominant. However, since
galaxy-wide processes such as galaxy mergers can trigger gas flows
into the bulge \citep{sand88,miho94}, it is conceivable that BHs could
regulate their growth on the scale of the entire galaxy
\citep[$\sim10$~kpc; ][]{haeh98,fabi99,wyit03} or even on that of the
DM haloes hosting the galaxies \citep[$\sim10^2$~kpc;
][]{silk98,ferr02}. Finally, it is possible, perhaps even likely, that
self-regulation takes place simultaneously on multiple scales. For
example, frequent, short, Eddington-limited outbursts may be able to
regulate the inflow of gas on the scale of the bulge averaged over
much longer time scales.

In this paper we investigate, using self-consistent simulations of the
co-evolution of the BH and galaxy populations, on what scale the
self-regulation of BHs takes place.  In Sec.~\ref{sec:method} we
describe the numerical techniques and simulation set employed in this
study.  In Sec.~\ref{sec:results} we demonstrate that BH
self-regulation takes place on the scale of the DM halo, and that the
BH mass is determined by the binding energy of the DM halo rather than
by the stellar mass of the host galaxy. Throughout we assume a flat
$\Lambda$CDM cosmology with the cosmological parameters:
$\{\Omega_{\rm m},\Omega_{\rm b},\Omega_\Lambda,\sigma_8,n_{\rm
  s},h\}=\{0.238,0.0418,0.762,0.74,0.951,0.73\}$, as determined from
the WMAP 3-year data \citep{sper07}.

\section{Numerical Methods}
\label{sec:method}
We have carried out a set of cosmological simulations using Smoothed
Particle Hydrodynamics (SPH).  We employ a significantly extended
version of the parallel PMTree-SPH code {\sc gadget iii} \citep[last
  described in ][]{spri05b}, a Lagrangian code used to calculate
gravitational and hydrodynamic forces on a particle by particle basis.
The initial particle positions and velocities are set at $z=127$ using
the Zeldovich approximation to linearly evolve positions from an
initially glass-like state.

In addition to hydrodynamic forces, we treat star formation, supernova
feedback, radiative cooling, chemodynamics and black hole accretion
and feedback, as described in \citet{scha08}, \citet{dall08},
\citet{wier08}, \citet{wier09} and \citet{boot09} (hereafter BS09)
respectively. For clarity we summarize here the essential features of
the BH model, which is itself a substantially modified version of that
introduced by \citet{spri05}.

\subsection{The black hole model}
Seed BHs of mass $m_{{\rm seed}}=10^{-3}m_{{\rm g}}$ -- where $m_{{\rm
    g}}$ is the simulation gas particle mass -- are placed into every
DM halo that contains more than 100 DM particles and does not already
contain a BH particle.  Haloes are identified by regularly running a
friends-of-friends group finder on-the-fly during the simulation.
After forming, BHs grow by two processes: accretion of ambient gas and
mergers.  Gas accretion occurs at the minimum of the Eddington rate,
$\dot{m}_{{\rm Edd}}=4\pi Gm_{{\rm BH}} m_{\rm p}/\epsr \sigma_{{\rm
    T}} c$ and $\dot{m}_{{\rm accr}}=\alpha4\pi G^2 m_{{\rm BH}}^2
\rho/(c_{s}^2+v^2)^{3/2}$, where $m_{\rm p}$ is the proton mass,
$\sigma_{\rm T}$ is the Thomson cross-section, $c$ is the speed of
light, $c_{s}$ and $\rho$ are the sound speed and density of the local
medium, $v$ is the velocity of the BH relative to the ambient medium,
and $\alpha$ is a dimensionless efficiency parameter.  The parameter
$\alpha$, which was set to 100 by \citet{spri05}, accounts for the
fact that our simulations possess neither the necessary resolution nor
the physics to accurately model accretion onto a BH on small scales.
Note that for $\alpha=1$ this accretion rate reduces to the so called
Bondi-Hoyle \citep{bond44} rate.

As long as we resolve the scales and physics relevant to Bondi-Hoyle
accretion, we could set $\alpha=1$.  If a simulation resolves the
Jeans scales in the accreting gas then it will also resolve the scales
relevant for Bondi-Hoyle accretion onto any BH larger than the
simulation mass resolution (BS09).  We therefore generally set
$\alpha$ equal to unity.  However, this argument breaks down in the
presence of a multi-phase interstellar medium, because our simulations
do not resolve the properties of the cold, molecular phase, and as
such the accretion rate may be orders of magnitude higher than the
Bondi-Hoyle rate predicted by our simulations for star-forming gas.
We therefore use a power-law scaling of the accretion efficiency such
that $\alpha=(n_{{\rm H}}/n_{{\rm H}}^*)^\beta$ in star-forming gas,
where $n_{{\rm H}}^*=0.1\,{\rm cm}^{-3}$ is the critical density for
the formation of a cold, star-forming gas phase.  The parameter
$\beta$ is a free parameter in our simulations.  We set $\beta=2$, but
note that the results shown here are insensitive to changes in this
parameter when $\beta \ga 2$ (see BS09), because in that case the
growth of the BHs is limited by feedback.

Energy feedback is implemented by allowing BHs to inject a fixed
fraction of the rest mass energy of the gas they accrete into the
surrounding medium.  The energy deposition rate is given by
\begin{equation}
\dot{E}=\epsf\epsr\dot{m}_{\rm accr}c^2=\frac{\epsf\epsr}{1-\epsr}\dot{m}_{\rm
BH}c^2\,,
\label{eq:edot}
\end{equation}
where $\epsr$ is the radiative efficiency of the BH, $\dot{m}_{\rm
  accr}$ is the rate at which the BH is accreting gas, and
$\dot{m}_{\rm BH}$ is the rate of BH mass growth.

We set $\epsr$ to be 0.1, the mean value for radiatively efficient
accretion onto a Schwarzschild BH \citep{shak73}. We vary $\epsf$ but
use $\epsf=0.15$ as our fiducial value.  It was shown in BS09 that,
for $\epsf=0.15$, simulations identical to these reproduce the
observed redshift zero $\mbh-\ms$ and $\mbh-\sigma$ relations, where
$\sigma$ is the one-dimensional velocity dispersion of the stars and
$\ms$ is the galaxy stellar mass.  Energy is returned to the
surroundings of the BH \lq thermally\rq, that is, by increasing the
temperature of $N_{\rm heat}$ of the BH's neighbouring SPH particles
by at least $\Delta T_{\rm min}$.  A BH performs no heating until it
has built up enough of an energy reservoir to heat by this amount.
The use of an energy reservoir is necessary in these simulations as
otherwise gas will be able to radiate away the energy every timestep.
Imposing a minimum temperature increase ensures that the radiative
cooling time is sufficiently long for the feedback to be effective.
In our fiducial model we set $N_{\rm heat}=1$ and $\Delta T_{\rm
  min}=10^8$\,K but the results are insensitive to the exact values of
these parameters (see BS09).

\subsection{The simulation set}
The simulations employed in the current work use cubic boxes of size
12.5 and 50 comoving Mpc/$h$ and assume periodic boundary
conditions. Each simulation contains either $128^3$ or $256^3$
particles of both gas and collisionless cold DM.  Comoving
gravitational softenings are set to $1/25$ of the mean interparticle
separation down to $z=2.91$, below which we switch to a fixed proper
scale.  The 12.5~Mpc/$h$ (50~Mpc/$h$) boxes are evolved as far as
redshift two (zero).  The numerical parameters of the simulations used
in this study are summarized in Table~\ref{tab:sims}.  All results
presented in this letter are derived from the 50.0~Mpc/$h$, $256^3$
particle simulations, with the other box sizes and particle numbers
employed to demonstrate numerical convergence.

\begin{table}\begin{center}
\caption{Numerical parameters of the simulations.  From left to right:
  Simulation identifier, comoving box size (Mpc/$h$), number of both
  gas and DM particles, final redshift, gas particle mass
  ($10^7\,\msun/h$), DM particle mass ($10^7\,\msun/h$), maximum
  physical gravitational softening (kpc/$h$).  Each simulation was run
  multiple times using different values of $\epsf$.}
\begin{tabular}{rrrrrrr}
Name & L & $n_{{\rm part}}$ & $z_{{\rm f}}$ & $m_{{\rm g}}$ & $m_{{\rm DM}}$ & $\epsilon_{{\rm max,phys}}$\\
\hline
\emph{L050N256} & 50.0 & 256$^3$ & 0 & $8.7$ & $41.0$ & 2.0\\
\emph{L050N128} & 50.0 & 128$^3$ & 0 & $69.6$ & $328.0$ & 4.0\\
\emph{L012N256} & 12.5 & 256$^3$ & 2 & $0.1$ & $0.6$ & 0.5\\
\end{tabular}
\label{tab:sims}
\end{center}\end{table}

\vspace{-0.4cm}
\section{Results and Discussion}
\label{sec:results}
It is instructive to first consider under what conditions BHs can
regulate their growth. To regulate its growth on a mass scale $\msr$,
a BH of mass $\mbh$ must be able to inject energy (or momentum) at a
rate that is sufficient to counteract the force of gravity on the
scale $\msr$, averaged over the dynamical time associated with this
scale. The mass $\msr$ could, for example, correspond to that of the
BH, the stellar bulge, or the dark matter (DM) halo. If the BH cannot
inject energy sufficiently rapidly, then gravity will win and its mass
will increase. Provided that the maximum rate at which it can inject
energy increases with $\mbh$ (as is for example the case for
Bondi-Hoyle and Eddington-limited accretion with a constant radiative
efficiency) and provided that this rate increases sufficiently rapidly
to counteract the growth of $\msr$, the BH will ultimately reach the
critical mass $m_{\rm BH,crit}(\msr)$ required to halt the inflow on
the scale $\msr$.  If, on the other hand, $\mbh \gg m_{\rm
  BH,crit}(\msr)$, then the BH will quickly quench the accretion flow
and its mass will consequently remain nearly unchanged. The BH will in
that case return to the equilibrium value $m_{\rm BH,crit}(\msr)$ on
the time scale which characterises the growth of $\msr$.

If the BH regulates its growth on the mass scale $\msr$ and if
$\mbh \ll \msr$, then the critical rate of energy injection required
for self-regulation is independent of the mass of the BH. It then
follows from Eq.~\ref{eq:edot} that $\dot{m}_{\rm BH}\propto
\epsf^{-1}$, which implies
\begin{equation}
\left (\mbh - \mseed\right ) \propto \epsf^{-1}\,,
\label{eq:prop}
\end{equation}
where $\mseed$ is the initial mass of the BH. Hence, if the
self-gravity of BHs is negligible on the maximum scale on which
they regulate their growth and if $\mbh \gg \mseed$, then
we expect $\mbh\propto
\epsf^{-1}$. 

\begin{figure}
\begin{center}
\includegraphics[width=8.3cm,clip]{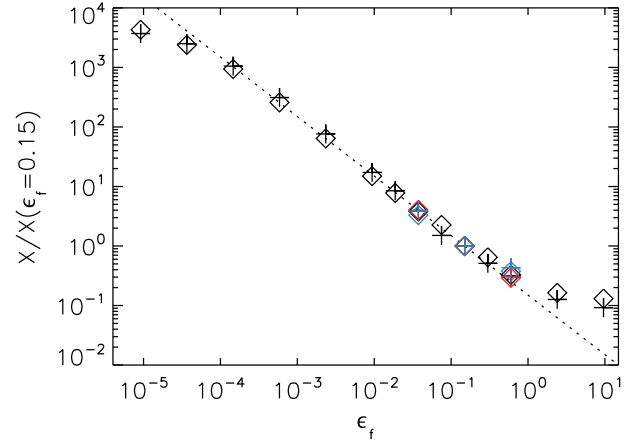}
\end{center}
\vspace{-0.4cm}
\caption{\label{fig:massscale}Predicted redshift zero global BH mass
  density (black diamonds) and normalization of the $\mbh-\sigma$
  relation (black plus signs) as a function of the assumed efficiency
  of BH feedback, $\epsf$. Both quantities are normalized to their
  values in the simulation with $\epsf=0.15$, which reproduces the
  observed relations between the mass of the BH and properties of the
  stellar bulge.  Each point represents a different simulation. For
  $10^{-4}<\epsf<1$ all data points track the dotted black line, which
  is a power-law with index minus one. This implies that in this
  regime BH mass is inversely proportional to $\mbh$, and thus that
  the BHs inject energy into their surroundings at a rate that is
  independent of $\epsf$, as expected for self-regulated growth on
  scales that are sufficiently large for the gravity of the BH to be
  unimportant. The red data points show results from simulations with
  a mass resolution that is 8 times worse than the fiducial
  simulation. The blue data points correspond to simulations with 64
  times better resolution than our fiducial resolution, but show
  results for redshift 2 rather than zero. The agreement between the
  black, red and blue points confirms numerical convergence and
  demonstrates that the BHs are already self-regulating at redshift
  2.}
\vspace{-0.4cm}
\end{figure}

\begin{figure*}
\begin{center}
\includegraphics[width=16.6cm,clip]{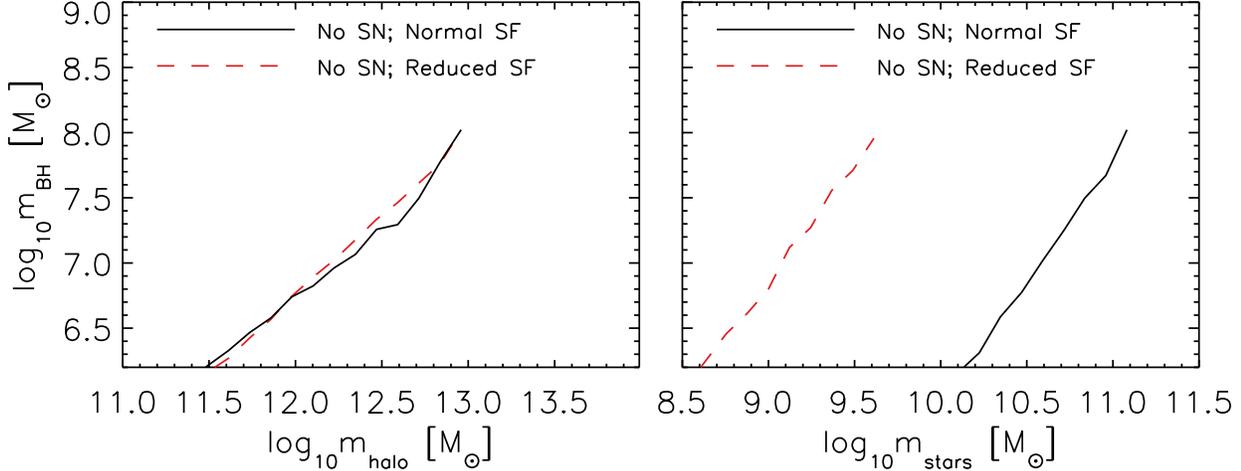}
\end{center}
\vspace{-0.8cm}
\caption{\label{fig:changing_baryons}Median $\mbh-\mhalo$ (\emph{left
    panel}) and $\mbh-\ms$ (\emph{right panel}) relations for all BHs
  more massive than $10m_{\rm seed}$. The black curves correspond to a
  simulation using our fiducial star formation law and the red, dashed
  curves show the result for a run in which the star formation
  efficiency was decreased by a factor of 100.  In order to isolate
  the effect of stellar mass, we turned off supernova feedback in both
  runs.  The BH scaling relations therefore differ somewhat from those
  predicted by our fiducial model, which does include supernova
  feedback.  Baryons dominate the gravitational potential in the
  central regions of the galaxy when we use our fiducial star
  formation law, but DM dominates everywhere in the run with the
  reduced star formation efficiency.  While the $\mbh-\ms$ relation is
  strongly affected by the change in the star formation efficiency,
  the relation between BH and halo mass remains invariant. This
  demonstrates that the BH mass is insensitive to the mass
  distribution on scales where the stellar mass dominates, and must
  instead be determined by the mass distribution on larger ($\gg
  10\,{\rm kpc}$) scales.}
\vspace{-0.4cm}
\end{figure*}

The black diamonds plotted in Fig.~\ref{fig:massscale} show the
predicted global mass density in BHs at redshift zero as a function of
$\epsf$, the efficiency with which BHs couple energy into the ISM,
normalised to the density obtained for $\epsf=0.15$. Similarly, the
black plus signs indicate the normalisation of the $\mbh-\sigma$
relation divided by that for the $\epsf=0.15$ run.  The feedback
efficiency, $\epsf$, is varied, in factors of 4, from
$\epsf=9.2\times10^{-6}$ to $\epsf=9.6$, which implies that the
fraction of the accreted rest mass energy that is injected
(i.e.\ $\epsr\epsf$) varies from $9.2\times 10^{-7}$ to 0.96.  BH mass
is clearly inversely proportional to the assumed feedback efficiency
for $10^{-4}<\epsf< 1$.  For $\epsf>1$ the trend breaks down because
the BH masses remain similar to the assumed seed mass, in accord with
Eq.~\ref{eq:prop}. If we had used a lower seed mass, then the trend
would have extended to greater values of $\epsf$. The deviation from
inverse proportionality that sets in below $\epsf =10^{-4}$ is more
interesting. Such low values yield BH masses that are more than
$0.15/10^{-4}\sim 10^{3}$ times greater than observed, in which case
they are no longer negligible compared to the masses of their host
galaxies. In that case the critical rate of energy deposition will no
longer be independent of $\mbh$ and we do not expect Eq.~\ref{eq:prop}
to hold.

We have thus confirmed that feedback enables BHs to regulate their
growth. Moreover, we demonstrated that this self-regulation takes
places on scales over which the gravitational influence of the BHs is
negligible, provided that the fraction of the accreted rest mass
energy that is coupled back into the interstellar medium is $\gtrsim
10^{-5}$.  

To test whether it is the stellar or the dark matter distribution that
determines the mass of BHs, we compare the BH masses in two
simulations that are identical except for the assumed efficiency of
star formation.  One uses our fiducial star formation law, but in the
other simulation we reduced its amplitude by a factor of 100, making
the gas consumption time scale much longer than the age of the
Universe. Because changing the amount of stars would imply changing
the rate of injection of supernova energy, which could affect the
efficiency of BH feedback, we neglected feedback from star formation
in both runs. In the simulation with \lq normal\rq\ star formation the
central regions of the galaxies are dominated gravitationally by the
baryonic component of the galaxy, whereas in the simulation with
reduced star formation the DM dominates everywhere.
Fig.~\ref{fig:changing_baryons} shows the $\mbh-\mhalo$ and $\mbh-\ms$
relations at redshift 0. While the two runs produce nearly identical
BH masses for a fixed halo mass, the $\mbh-\ms$ relation is shifted to
lower stellar masses by more than an order of magnitude in the model
with reduced star formation.  The insensitivity of the relation
between $\mbh$ and $\mhalo$ to the assumed star formation efficiency
demonstrates that the BH mass is not set by the gravitational
potential on the scale of the galaxy.  We have verified that the same
result holds at redshift two for the simulations with 64 times better
mass resolution.  Clearly, stellar mass does not significantly
influence the relation between the mass of the BH and that of its host
halo. This implies that BH self-regulation occurs on the scale of DM
haloes.

If the rate by which the BHs inject energy is independent of the
assumed feedback efficiency, then we expect the same to be true for
the factor by which BH feedback suppresses star formation. This is
confirmed by comparison of the global SFRs in runs with different
values of $\epsf$ (see Fig.~6 of BS09).

\begin{figure}
\begin{center}
\includegraphics[width=8.3cm,clip]{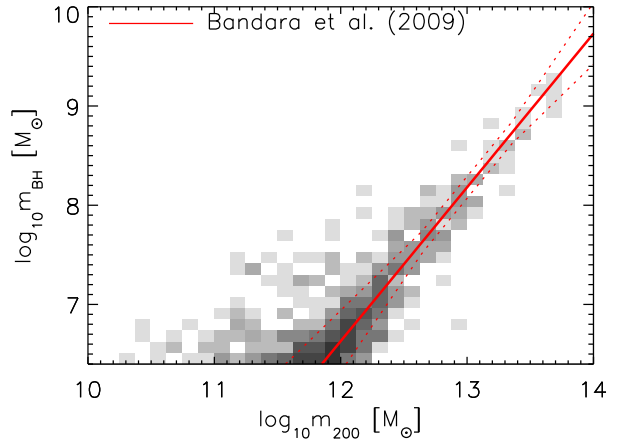}
\end{center}
\vspace{-0.4cm}
\caption{\label{fig:mhalombh}The relation between BH mass and DM halo
  mass for all BHs that belong to central galaxies and have masses
  greater than $10m_{\rm seed}$. The DM halo mass, $m_{200}$, is
  defined as the mass enclosed within a sphere, centred on the
  potential minimum of the DM halo, that has a mean internal density
  of 200 times the critical density of the Universe.  The grey pixels
  show the results from our fiducial simulation ($\epsf=0.15$), with
  the colour of each pixel set by the logarithm of the number of BHs
  in that pixel.  The solid, red line shows the observational
  determination of the $\mbh-\mhalo$ relation (Bandara et al. 2009)
  and has a slope of 1.55. The dotted, red lines show the 1$\sigma$
  errors on the observations.  The simulation agrees very well with
  the observed relation.  The value of the slope and the scatter
  (which correlates with the concentration of the DM halo) suggest
  that the halo binding energy, rather than mass, determines the
  masses of BHs.}
\vspace{-0.4cm}
\end{figure}

Fig.~\ref{fig:mhalombh} compares the predicted
$\log_{10}\mbh-\log_{10}\mhalo$ relation with observation
\citep{band09}.  The agreement is striking. The slope and
normalization of the observed
$\log_{10}(\mbh/\msun)-\log_{10}(\mhalo/10^{13}\,\msun)$ relation are
$1.55\pm0.31$ and $8.18\pm0.11$ respectively, whereas the simulation
predicts $1.55\pm0.05$ and $8.01\pm0.04$.  Note that the simulation
was only tuned to match the normalization of the relations between
$\mbh$ and the galaxy stellar properties.

If the energy injected by a BH is proportional to the halo
gravitational binding energy, then, for isothermal models
\citep{silk98}, $\mbh\propto\mhalo^{5/3}$.  Here we extend these
models to the more realistic universal halo density profile
\citep{nava97}, whose shape is specified by a concentration parameter,
$c$ (we assumed $c\propto v_{\rm max}^2/v_{\rm v}^2$, where $v_{\rm
  max}$ and $v_{{\rm v}}$ are the maximum halo circular velocity and
the circular velocity at the virial radius respectively).  It is known
that concentration decreases with increasing halo mass,
$c\propto\mhalo^{-0.1}$ \citep{bulo01,duff08}, which then affects BH
mass through the dependence of halo binding energy on concentration.
If the total energy injected by a BH of a given mass is proportional
to the energy required to unbind gas from a DM halo \citep{loka01} out
to some fraction of the virial radius, $\re/r_{\rm v}$ then
\begin{align}
\mbh\propto&\Bigg(\frac{c}{\big(\ln(1+c)-c/(1+c)\big)^2}\Bigg)\times \nonumber\\
&\Bigg(1-\frac{1}{(1+c\frac{\re}{r_{\rm v}})^2}-\frac{2\ln(1+c\frac{\re}{r_{\rm v}})}{1+c\frac{\re}{r_{\rm v}}}\Bigg)m_{\rm v}^{5/3}\,.
\end{align}
Inserting $c\propto m_{\rm v}^{-0.1}$ and computing the logarithmic
derivative with respect to $m_{\rm v}$ in the mass range
$10^{10}\,\msun < m_{\rm v} < 10^{14}\,\msun$, we find that the slope
is a weak function of $\re/r_{\rm v}$ that varies from 1.50 at
$\re=10^{-1}r_{\rm v}$ to 1.61 at $\re=r_{\rm v}$.  The close match
between theory, simulation and observation suggests that the halo
binding energy, rather than halo mass, determines the mass of the BH.

The residuals from the $\mbh-\mhalo$ relation ($\Delta\log_{10}\mbh$)
are correlated with halo concentration (Spearman rank correlation
coefficient $\rho=0.29$, probability of significance $P= 0.9998$) as
would be expected if $\mbh$ is sensitive to the halo binding energy.
The residuals are also correlated with galaxy stellar mass, though
much less strongly ($\rho=0.09$; $P=0.96$).  Taken together, these
correlations tell us that, at a given halo mass, galaxies with BHs
more massive than the average will also contain a larger than average
amount of stars, and are hosted by more concentrated haloes. This
suggests that the galaxy stellar mass is also determined by the halo
binding energy.  Thus, outliers in the $\mbh-\mhalo$ relation may
still lie close to the mean $\mbh-\ms$ relation.  Furthermore, higher
concentrations imply earlier formation times and spheroidal components
do indeed typically host old stellar populations.

In addition to the \lq quasar mode\rq\ of feedback discussed in this
work, it has recently become clear that a second \lq radio
mode\rq\ may be required to quench cooling flows in galaxy groups and
clusters \citep[see e.g.][for a review]{catt09}.  Although we do not
explicitly include a \lq radio mode\rq\ in the current work, the AGN
feedback prescription explored here is capable of suppressing cooling
flows, at least on group scales, providing excellent matches to
observed group density and temperature profiles as well as galaxy
stellar masses and age distributions \citep{mcca09}. It is known that
BHs obtain most of their mass in the \lq quasar
mode\rq\ \citep{solt82} so any discussion of what detemines the masses
of BHs must focus primarily on this mode of accretion.  Finally, the
ability of a BH to quench cooling flows in the \lq radio mode\rq\ is
expected to be closely related the virial properties of the hot halo
\citep{catt09} and would therefore provide an additional link between
BHs and DM haloes over and above what we discuss here and so serve to
make any fundamental connection between BH mass and the properties of
the DM halo even stronger.

We conclude that our simulation results suggest that in order to
effectively halt BH (and galaxy) growth, gas must not return to the
galaxy on a short timescale. This requires that the BH injects enough
energy to eject gas out to scales where the DM halo potential is
dominant. The mass of the BH is therefore determined primarily by the
mass of the DM halo with a secondary dependence on halo concentration,
of the form that would be expected if the BH mass were controlled by
the halo binding energy.  The tight correlation between $\mbh$ and
$\ms$ is then a consequence of the more fundamental relations between
halo binding energy and both $\mbh$ and $\ms$.

\vspace{-0.4cm}
\section*{Acknowledgments}
The authors thank Marijn Franx and the referee, Andrea Cattaneo, for
useful discussions and suggestions.  The simulations presented here
were run on the Cosmology Machine at the Institute for Computational
Cosmology in Durham as part of the Virgo Consortium research
programme.  This work was supported by an NWO Vidi grant.

\label{lastpage}

\end{document}